\documentclass[usenatbib]{mn2e}

\usepackage{graphicx}

\title[Observations of the Cepheid $\ell$~Car]{Observations of the pulsation of the Cepheid
$\ell$~Car with the Sydney University Stellar Interferometer}

\author[J. Davis et al.]{J. Davis,\thanks{E-mail: j.davis@physics.usyd.edu.au}
A.P. Jacob, J.G. Robertson, M.J. Ireland, J.R. North, W.J. Tango \newauthor and P.G. Tuthill\\
Sydney Institute for Astronomy, School of Physics, University of Sydney, NSW 2006, Australia}

\begin{document}

\maketitle

\begin{abstract}
Observations of the southern Cepheid $\ell$~Car to yield the mean
angular diameter and angular pulsation amplitude have been made
with the Sydney University Stellar Interferometer (SUSI) at a wavelength
of 696\,nm.  The resulting mean limb-darkened angular diameter is 2.990$\pm$0.017\,mas (i.e. $\pm$0.6\,per cent)
with a maximum-to-minimum amplitude of 0.560$\pm$0.018\,mas corresponding to 18.7$\pm$0.6\,per cent
in the mean stellar diameter.  Careful attention has been paid to uncertainties, including those in
measurements, in the adopted calibrator angular diameters, in the projected values of
visibility squared at zero baseline, and to systematic effects.  No evidence was found for a circumstellar
envelope at 696\,nm.  The interferometric results
have been combined with radial displacements of the stellar atmosphere derived from selected
radial velocity data taken from the literature to determine the distance and mean diameter of $\ell$~Car.
The distance is determined to be 525$\pm$26\,pc and the mean radius 169$\pm$8\,$R_{\sun}$.  Comparison
with published values for the distance and mean radius show excellent agreement, particularly when
a common scaling factor from observed radial velocity to pulsation velocity of the stellar atmosphere (the
$p$-factor) is used.
\end{abstract}


\section{Introduction}\label{sec:intro}

The accurate determination of the zero point of the Cepheid
Period-Luminosity relation is an important step in the refinement
of the extragalactic distance scale.  The combination of
interferometric and spectroscopic data, as discussed by
\citet{99jd}, is an approach to this task that is now feasible.
The first demonstration of the combination of interferometric and
spectroscopic data to determine both the mean
diameter and distance of a Cepheid was made by \citet{00lane}
based on interferometric observations with the Palomar Testbed Interferometer (PTI)
of $\zeta$~Gem.  \citet*{02lane} subsequently applied the technique
to $\eta$~Aql and, with additional data, to $\zeta$~Gem, again based on
PTI observations.  \citet{04K1} have reported the measurement of the angular
pulsations of seven Galactic Cepheids made with the European
Southern Observatory's Very Large Telescope Interferometer (VLTI) and,
in combination with spectroscopic data, have determined the distances to
the Cepheids.

One of the prime programmes for which the Sydney University
Stellar Interferometer (SUSI) \citep{99susi1} was developed was
the measurement of the mean diameters and angular pulsation
amplitudes of Cepheids but, in its initial configuration, it
lacked the required sensitivity. The recent commissioning of a red
beam-combination system in SUSI \citep{07susi} has resulted in a
significant increase in sensitivity.  In this paper we report SUSI
measurements of the mean angular diameter and angular pulsation
amplitude of $\ell$~Car, one of the seven Cepheids measured with
the VLTI by \citet{04K1}.  A preliminary analysis was presented at
a European Southern Observatory symposium in 2005 \citep{07eso} and
\citet{08apj} has discussed the SUSI Cepheid programme including an
analysis of a subset of the $\ell$~Car observations reported here.

Spectroscopic radial velocity measurements from the literature
have been combined to produce a radial velocity versus pulsation
phase curve and, after scaling with a fixed $p$-factor as discussed
in Section~\ref{sec:p-factor}, this has been integrated to give the radial
displacement of the Cepheid surface as a function of pulsation
phase.  The radial displacement and limb-darkened angular diameter
data have been combined to determine the distance and mean radius
of $\ell$~Car.

The resulting distance and mean radius are compared with values in the
literature determined by the same technique and by the infrared surface brightness
method and excellent agreement is found, particularly when all values are scaled
to a common $p$-factor.

\section{The Pulsation Phase} \label{sec:ephem}

Although it might seem premature to discuss the pulsation phase
prior to discussing the observations it is necessary to ensure
that both interferometric and spectroscopic phases are
synchronised since the epochs of the interferometric and spectroscopic
observations do not overlap.  A problem in the case of $\ell$~Car is
the fact that there is evidence that its period has undergone changes.
\citet{89szab} reported that the period of $\ell$~Car changed from
35.5318\,days prior to JD2440000 to 35.5513\,days.  \citet{92shob}
refined the Szabados period, with additional photometry in 1990,
to 35.5443\,days and \citet{04bandt}, based on photometry in 2002,
give a value of 35.5572\,days.  \citet{97tetal} have listed values
back to 1901 and found in their own spectroscopic study of
$\ell$~Car from early 1991 to late 1996 that an increase in the
period was evident although the uncertainties were large.

The approach that we have adopted is as follows.  Since the
largest body of radial velocity data available to us is the
combined Mount Stromlo Observatory (MSO) and Mount John University
Observatory (MJUO) observations analysed by \citet{97tetal} and
their phases are effectively based on the \cite{92shob} ephemeris,
albeit with small adjustments to bring the early MJUO data into
line with the MSO data, we have adopted the \citet{92shob}
ephemeris for the analysis of both interferometric and additional
spectroscopic data.  The adopted period is
35.5443$\pm$0.0006\,days with 2447880.81$\pm$0.10 the zero point of
maximum light in heliocentric Julian Date.  The phases of individual
data points have been computed from the Julian Date of observation
for both interferometric and spectroscopic data.

Because of the variations in period it is found that there are
small phase shifts between data sets taken at different epochs.
The change in phase across any given data set
as a result of adopting the Shobbrook period rather than for
example, that of \citet{04bandt}, is negligible, and a small phase
shift of the whole data set to bring it into alignment with the
\citet{97tetal} data is justified.  Details of the alignment
of the different sets of data will be discussed when combining
data and, in particular, spectroscopic data in
Section~\ref{sec:spec}.

\section{The Interferometric Observations}\label{sec:obs}

Measurements of the squared fringe visibility $V^{2}$ were made
with SUSI \citep{99susi1} using the red beam-combination system,
which employs the fringe-scanning technique \citep{07susi}. This
beam-combination system uses matched filters with a central
wavelength of 700\,nm and spectral bandwidth of 80\,nm.  The
observations to determine the angular diameter of $\ell$~Car were
made with a baseline of 40\,m with additional measurements at 5\,m
to enable the zero baseline value of $V^{2}$ to be checked.

\subsection{Calibrators} \label{sec:cals}

Calibrators were selected as close in the sky as possible to
$\ell$~Car with the additional requirement of being minimally
resolved. The limiting sensitivity of SUSI at 700\,nm is $\sim$+5
which limited the choice of calibrators and compromise was
necessary. The calibrators used are listed in Table~\ref{tab:cals}
with their spectral types, adopted uniform-disk angular diameters,
and angular distances from $\ell$~Car.  Common calibrators were
used throughout to eliminate the potential influence of calibrator
diameters on the pulsation curve.  The effective wavelength
of observations of all three calibrators has been estimated to be
695.0$\pm$2.0\,nm following a similar analysis to that described
by \citet{07susi}.

The uniform-disk angular diameters have been determined from
measurements made with the Narrabri Stellar Intensity
Interferometer (NSII) \citep*{74nsii} and with the Mark III Optical
Interferometer (Mark III) \citep{03mk3}.  In the case of
$\beta$~Car the value measured with the NSII has been adopted
after correction from the limb-darkened angular diameter to the
uniform-disk angular diameter using the appropriate correction
factor for 695\,nm interpolated from the tabulation of \citet*{00dtb}.  In the absence
of measured angular diameters for $\iota$ and s~Car, limb-darkened
angular diameters have been determined by interpolation in a plot of NSII
and Mark III limb-darkened angular diameters for unreddened visual magnitude $V_{0}=0.0$ versus
($B$-$V$)$_{0}$ for ($B$-$V$)$_{0} < 0.6$.  The interpolated limb-darkened angular diameters
have been corrected to the $V_{0}$ magnitudes of the stars and
corrected for limb-darkening to give the uniform-disk angular
diameters listed in Table~\ref{tab:cals}, again using appropriate
correction factors interpolated from the tabulation of
\citet{00dtb}.  The uncertainties in the uniform-disk angular diameters have
been estimated from the scatter in the plot of limb-darkened angular diameters
versus ($B$-$V$)$_{0}$.

\begin{table}
  \caption{Calibrators used for the observations of $\ell$~Car.
  $m_{695}$ is the estimated magnitude at 695\,nm, $\theta_{\mathrm{UD}}$
  the adopted uniform disk angular diameter at 695\,nm and $\Omega$
  the angular distance of the calibrator from $\ell$~Car.}
  \label{tab:cals}
  \begin{tabular}{cccccc}
  \hline
  HR & Star & Spectral & $m_{695}$ & $\theta_{\mathrm{UD}}$ & $\Omega$  \\
  & & Type & & (mas) & (deg)  \\
  \noalign{\smallskip}\hline\noalign{\smallskip}
  3685 & $\beta$ Car  & A2 IV  & 1.6 & 1.54$\pm$0.07 &  7.9  \\
  3699 & $\iota$ Car  & A8 Ib  & 2.0 & 1.55$\pm$0.12 &  4.7  \\
  4114 & s Car        & F2 II  & 3.5 & 0.90$\pm$0.07 &  6.5  \\
  \hline
  \end{tabular}
\end{table}

\subsection{The Observations} \label{sec:obs2}

Observations of calibrators and $\ell$~Car were alternated in each
observing session so that every $\ell$~Car observation was
bracketed by observations of calibrators. Each observation
consisted of a set of 1000 scans each 140\,$\mu$m long and
consisting of 1024 by 0.2\,ms samples. Each scan set was followed
by photometric and dark scans.  One complete observation of
$\ell$~Car, including the bracketing calibrators, took a total of
$\sim$18 minutes.

Observations were made with a baseline of 40\,m on 31 nights for
$\ell$~Car between 2 March 2004 and 24 May 2007.  Initially all
31 nights were included in the analysis but four nights were
subsequently rejected when their values were found to lie more than four standard
deviations from the fit to limb-darkened angular
diameter versus radial displacement of the stellar surface (Section~\ref{sec:ldadvrd})
and the analysis was then repeated without them.  The maximum deviation from
the fit after rejection of the four points was $<3.1\sigma$.  Examination of the data
for three of the four rejected nights revealed that they were obtained during a
short period in late March-early April 2004 when there was
significant leakage of the metrology laser light into the signal beams.
On the fourth rejected night, the data were poor, showing large scatter with
only two out of the six points lying within one standard deviation of the
mean for the night.

The projected baseline (i.e. the effective baseline of
an observation) was less than 40\,m due to the southerly declinations of
$\ell$~Car and its calibrators.

Observations were also made with a 5\,m baseline, close to the
phases of maximum and minimum angular diameter, on 13~March and
5~May 2007.  These will be discussed in Section~\ref{sec:5m}.

\section{Analysis of Interferometry}\label{sec:analysis}

The initial analysis of the fringe scans for $\ell$~Car and the
calibrators was carried out in the SUSI software ``pipeline''
\citep{07susi} which outputs the raw and seeing corrected squared
visibility amplitudes ($V^{2}$), the Julian Date, projected
baseline, hour angle, fluxes etc. for each set of scans.  For the
early observations up to 7 April 2004 full sets of photometry
files were not recorded.  In these cases photometry files were
duplicated to enable the scans to be processed in the pipeline and
it is noted that analysis has shown that this procedure has a
negligible effect on the resulting values of $V^{2}$.  For each
night the output file from the pipeline was imported into an Excel
spreadsheet for examination of the data and for further analysis.

For each set of calibrator scans a Transfer Function ($T$) was
calculated.  $T$ is given by
 \begin{equation}
T = \frac{V^{2}_{\mathrm{obs}}}{V^{2}_{\mathrm{exp}}}
 \end{equation}
where $V^{2}_{\mathrm{obs}}$ is the observed, seeing corrected
value of $V^{2}$ and $V^{2}_{\mathrm{exp}}$ is the expected value
of $V^{2}$ calculated from the uniform-disk angular diameter,
effective wavelength and projected baseline.

The values of $V^{2}_{\mathrm{obs}}$ for $\ell$~Car were then
multiplied by the weighted mean value of $T$ for the two
calibrators bracketing the $\ell$~Car observation to give the `true'
calibrated value of $V^{2}$ for $\ell$~Car.  That is

\begin{equation}
V^{2}_{\mathrm{true}}(\ell~\mathrm{Car}) =
\frac{V^{2}_{\mathrm{obs}}(\ell~\mathrm{Car})}{\overline{T}}
\end{equation}

The mean value of $V^{2}_{\mathrm{true}}(\ell~\mathrm{Car})$ for
each night is listed in Table~\ref{tab:vals} with the date (Universal Time), the
mean Julian Date (JD), the mean projected baseline, the
calibrators used, and the number of scan sets for $\ell$~Car.  The uncertainty
in $V^{2}_{\mathrm{true}}(\ell~\mathrm{Car})$ takes into account the uncertainties in the
uniform-disk angular diameters of the calibrators although these are generally negligible
compared to the uncertainties in the values of $V^{2}_{\mathrm{obs}}$ for both
calibrators and $\ell$~Car.

\begin{table}
  \caption{Calibrated values of $V^{2}$.  The format for the date (Universal Time) is ddmmyy, $\overline{\mathrm{JD}}$ is the mean
  Julian Date of the observations minus 2450000, $\overline{b}$ is the mean projected
  baseline, and $\overline{V^{2}_{\mathrm{true}}(\ell~\mathrm{Car})}$ is the mean of N values of $V^{2}_{\mathrm{true}}(\ell~\mathrm{Car})$.
  Further details are given in the text.}
  \label{tab:vals}
  \begin{tabular}{cccc@{\hspace{3mm}}c@{\hspace{3mm}}c}
  \hline
  Date & $\overline{\mathrm{JD}}$ & $\overline{b}$ & $\overline{V^{2}_{\mathrm{true}}(\ell~\mathrm{Car})}$ & Calibrators & N \\
   & & (m) & & (Carinae) & \\
  \hline
  020304 & 3067.039 & 33.72 & 0.388$\pm$0.020 & $\beta$ \& $\iota$ & 8 \\
  070304 & 3072.020 & 33.41 & 0.385$\pm$0.010 & $\beta$ \& $\iota$ & 8 \\
  170304 & 3081.940 & 33.49 & 0.274$\pm$0.013 & $\beta$ \& $\iota$ & 3 \\
  180304 & 3083.028 & 33.69 & 0.269$\pm$0.004 & $\beta$ \& $\iota$ & 3 \\
  010404 & 3096.964 & 33.67 & 0.301$\pm$0.014 & $\beta$ \& $\iota$ & 12 \\
  140404 & 3109.958 & 33.59 & 0.349$\pm$0.006 & $\beta$, $\iota$ \& s & 7 \\
  160404 & 3112.026 & 32.53 & 0.341$\pm$0.005 & $\beta$, $\iota$ \& s & 3 \\
  170404 & 3112.976 & 33.42 & 0.334$\pm$0.016 & $\beta$, $\iota$ \& s & 6 \\
  180404 & 3113.934 & 33.76 & 0.294$\pm$0.009 & $\beta$, $\iota$ \& s & 6 \\
  200404 & 3115.940 & 33.74 & 0.278$\pm$0.005 & $\beta$, $\iota$ \& s & 7 \\
  210404 & 3116.976 & 33.26 & 0.286$\pm$0.005 & $\beta$, $\iota$ \& s & 6 \\
  220404 & 3117.900 & 33.82 & 0.255$\pm$0.009 & $\beta$, $\iota$ \& s & 6 \\
  240404 & 3119.933 & 33.65 & 0.284$\pm$0.010 & $\beta$, $\iota$ \& s & 5 \\
  300404 & 3125.918 & 33.69 & 0.278$\pm$0.018 & $\beta$, $\iota$ \& s & 3 \\
  040504 & 3129.899 & 33.65 & 0.261$\pm$0.011 & $\beta$, $\iota$ \& s & 7 \\
  080504 & 3133.958 & 32.52 & 0.326$\pm$0.013 & $\beta$, $\iota$ \& s & 4 \\
  120504 & 3137.933 & 32.93 & 0.397$\pm$0.014 & $\beta$, $\iota$ \& s & 9 \\
  180105 & 3389.187 & 33.66 & 0.388$\pm$0.015 & $\beta$, $\iota$ \& s & 9 \\
  030205 & 3405.120 & 33.67 & 0.241$\pm$0.006 & $\beta$, $\iota$ \& s & 13 \\
  040205 & 3406.117 & 33.70 & 0.241$\pm$0.004 & $\beta$, $\iota$ \& s & 12 \\
  050205 & 3407.095 & 33.75 & 0.239$\pm$0.008 & $\beta$, $\iota$ \& s & 8 \\
  060205 & 3408.070 & 33.59 & 0.213$\pm$0.007 & $\beta$, $\iota$ \& s & 3 \\
  070205 & 3409.144 & 33.61 & 0.233$\pm$0.004 & $\beta$, $\iota$ \& s & 9 \\
  100205 & 3412.117 & 33.62 & 0.258$\pm$0.007 & $\beta$, $\iota$ \& s & 7 \\
  110205 & 3413.150 & 33.52 & 0.256$\pm$0.004 & $\beta$, $\iota$ \& s & 5 \\
  120205 & 3414.140 & 33.56 & 0.295$\pm$0.005 & $\beta$, $\iota$ \& s & 6 \\
  240507 & 4244.919 & 32.48 & 0.391$\pm$0.005 & $\beta$ \& s & 7 \\
  \hline
  \end{tabular}
\end{table}

\section{The Angular Diameter} \label{sec:ad}

The uniform-disk angular diameter ($\theta_{\mathrm{UD}}$) for
each night was determined by fitting the equation

\begin{equation}
V^{2} = \left|\frac{2J_{1}(x)}{x}\right|^{2}
\label{eqn:bess}
\end{equation}
to the individual values of
$V^{2}_{\mathrm{true}}(\ell~\mathrm{Car})$ assuming that
$V^{2}_{\mathrm{true}}(\ell~\mathrm{Car})$ at zero baseline was
unity.  In equation (\ref{eqn:bess}) $J_{1}(x)$ is a Bessel
function and $x = \pi\,b\,\theta_{\mathrm{UD}}/\lambda$ where $b$ is
the projected baseline and $\lambda$ is the wavelength of the
observation.

The effective wavelength for $\ell$~Car will be a function of
pulsation phase as the star changes colour during its cycle.
Calculations of effective wavelength for supergiants following the
procedure described by \citet{07susi} give the mean effective
wavelength for $\ell$~Car as 696.2\,nm with a variation of
$\sim\pm$0.7\,nm, or $\pm$0.1\%, over a complete pulsation
cycle.  The systematic uncertainty in the calculated effective wavelength is
conservatively estimated to be $\pm$2.0\,nm and we therefore adopt
a fixed value of 696.2$\pm$2.0\,nm as the effective wavelength for
$\ell$~Car. Since the expected variation is significantly less
than the adopted uncertainty and, generally, small compared
with the observational uncertainties in the angular diameters, the
adoption of a fixed value is justified.

\subsection{The Zero Baseline Value of $V^{2}_{\mathrm{true}}(\ell~\mathrm{Car})$} \label{sec:5m}

It is important to check the validity of the assumption that
$V^{2}_{\mathrm{true}}(\ell~\mathrm{Car})$ is unity at zero baseline.  If
it differs from unity as a result of the presence of a companion or scattered
light from extended circumstellar matter, for example, it would translate to
errors in the angular diameters.  A large circumstellar envelope
(CSE) has been discovered around $\ell$~Car at $N$-band using long-baseline
interferometry by \citet{06ketal} and its signature was detected by
the same authors in their $K$-band observations.  CSEs have also been detected
in the $K$-band around Polaris and $\delta$~Cep by \citet{06metal} and around Y~Oph by
\citet{07metal}.  \citet{08nar} confirm a dominant absorption
component in H$\alpha$ for $\ell$~Car, whose velocity is constant and near
zero in the stellar rest frame, and they attribute this to a CSE.  These
detections of CSEs, particularly around $\ell$~Car, emphasise the importance of
checking the validity of our assumption.  In order to carry out this check,
observations were made at a baseline of 5\,m on 13 March
2007 at phase 0.023 and on 5 May 2007 at phase 0.514.  The
observations were therefore close to minimum and maximum angular
diameter respectively.  The observations were made after the
initial analysis of the 40\,m data had been completed so values
for the uniform-disk angular diameter at the two phases could be
interpolated from a plot of uniform-disk angular diameter versus
phase.  Given the angular diameter, and assuming
$V^{2}_{\mathrm{true}}(\ell~\mathrm{Car})$ to be unity at zero
baseline, the expected value of
$V^{2}_{\mathrm{true}}(\ell~\mathrm{Car})$ at the projected
baseline for each night was calculated.  If the assumption that
$V^{2}_{\mathrm{true}}(\ell~\mathrm{Car})$ is unity at zero
baseline is valid the observed and expected values will be in
agreement.  The results are listed in Table~\ref{tab:5m} and show
that at minimum angular diameter the observed and expected values
of $V^{2}_{\mathrm{true}}(\ell~\mathrm{Car})$ differ by
0.004$\pm$0.013 and at maximum angular diameter by
0.004$\pm$0.012. The assumption that
$V^{2}_{\mathrm{true}}(\ell~\mathrm{Car})$ is unity at zero
baseline is therefore justified and there is no evidence
of a significant effect of a CSE at 696\,nm.

\begin{table*}
  \caption{The results of the 5\,m observations near minimum and maximum values of the angular diameter
  of $\ell$~Car.  The format for the date is ddmmyy, $\overline{\mathrm{JD}}$ is the mean
  Julian Date of the observations minus 2450000, $\phi$ the mean pulsation phase for the
  observations, $\theta_{\mathrm{UD}}$ is the estimated uniform-disk angular diameter,
  $\overline{b}$ the mean baseline of the observations, $V^{2}_{\mathrm{exp}}$ the expected value of $V^{2}$ for the
  uniform-disk angular diameter, baseline and effective wavelength (696.2\,nm), $V^{2}_{\mathrm{obs}}$ is the mean value
  of the N observed values of  $V^{2}$, and ($V^{2}_{\mathrm{exp}}$-$V^{2}_{\mathrm{obs}}$) is the difference between
  the expected and observed values of $V^{2}$.  Further details are given in the text.}
  \label{tab:5m}
  \begin{tabular}{ccccccccc}
  \hline
   Date & $\overline{\mathrm{JD}}$ &  $\phi$ & $\theta_{\mathrm{UD}}$ & $\overline{b}$ & $V^{2}_{\mathrm{exp}}$ & $V^{2}_{\mathrm{obs}}$ & N & ($V^{2}_{\mathrm{exp}}$-$V^{2}_{\mathrm{obs}}$) \\
    &  &   & (mas) & (m) & & & \\
  \hline
  130307 & 4172.984 & 0.023 & 2.62$\pm$0.01 & 4.21 & 0.9855$\pm$0.0001 & 0.990$\pm$0.013
  & 5 & 0.004$\pm$0.013 \\
  050507 & 4225.968 & 0.514 & 3.13$\pm$0.01 & 4.08 & 0.9806$\pm$0.0001 &
  0.985$\pm$0.012 & 8 & 0.004$\pm$0.012 \\
  \hline
  \end{tabular}
\end{table*}

\subsection{The Uniform-Disk Angular Diameters} \label{sec:udads}

The uniform-disk angular diameters corresponding to the dates of
observation, together with the corresponding pulsation phases
determined using the ephemeris of \citet{92shob} as discussed in
Section~\ref{sec:ephem}, are listed in Table~\ref{tab:ads}.

\begin{table}
  \caption{The angular diameter of $\ell$~Car.
  $\overline{\mathrm{JD}}$ is the mean
  Julian Date of the observations minus 2450000 (see Table~\ref{tab:vals}), $\phi$ the mean pulsation phase for the
  observations, $\theta_{\mathrm{UD}}$ the uniform-disk angular diameter, Red. $\chi^{2}$ the reduced $\chi^{2}$ of the fit
  of equation~(\ref{eqn:bess}) to the values of $V^{2}_{\mathrm{true}}(\ell~\mathrm{Car})$, $\rho_{\mathrm{696}}$ the limb-darkening
  factor, and $\theta_{\mathrm{LD}}$ is the limb-darkened angular diameter.  Further details are given in the text.}
  \label{tab:ads}
  \begin{tabular}{c@{\hspace{3mm}}c@{\hspace{3mm}}c@{\hspace{3mm}}c@{\hspace{3mm}}c@{\hspace{3mm}}c}
  \hline
   $\overline{\mathrm{JD}}$ & $\phi$ & $\theta_{\mathrm{UD}}$ & Red.  & $\rho_{\mathrm{696}}$ & $\theta_{\mathrm{LD}}$ \\
   & & (mas) & $\chi^{2}$ & & (mas)  \\
  \hline
  3067.039 & 0.909 & 2.575$\pm$0.052 & 1.15 & 1.052 & 2.709$\pm$0.055 \\
  3072.020 & 0.049 & 2.568$\pm$0.026 & 0.73 & 1.049 & 2.694$\pm$0.027 \\
  3081.940 & 0.328 & 2.933$\pm$0.031 & 1.51 & 1.060 & 3.109$\pm$0.033 \\
  3083.028 & 0.359 & 2.933$\pm$0.053 & 0.02 & 1.060 & 3.109$\pm$0.056 \\
  3096.964 & 0.751 & 2.814$\pm$0.016 & 1.01 & 1.061 & 2.986$\pm$0.017 \\
  3109.958 & 0.116 & 2.678$\pm$0.020 & 0.61 & 1.051 & 2.815$\pm$0.021 \\
  3112.026 & 0.174 & 2.787$\pm$0.031 & 0.34 & 1.054 & 2.937$\pm$0.033 \\
  3112.976 & 0.201 & 2.740$\pm$0.031 & 1.90 & 1.055 & 2.891$\pm$0.033 \\
  3113.934 & 0.228 & 2.850$\pm$0.044 & 0.46 & 1.056 & 3.010$\pm$0.046 \\
  3115.940 & 0.284 & 2.900$\pm$0.016 & 0.82 & 1.058 & 3.068$\pm$0.017 \\
  3116.976 & 0.314 & 2.914$\pm$0.015 & 1.31 & 1.059 & 3.086$\pm$0.016 \\
  3117.900 & 0.340 & 2.986$\pm$0.018 & 2.28 & 1.060 & 3.165$\pm$0.019 \\
  3119.933 & 0.397 & 2.889$\pm$0.048 & 0.46 & 1.061 & 3.065$\pm$0.051 \\
  3125.918 & 0.565 & 2.897$\pm$0.033 & 3.61 & 1.063 & 3.080$\pm$0.035 \\
  3129.899 & 0.677 & 2.909$\pm$0.015 & 2.37 & 1.063 & 3.092$\pm$0.016 \\
  3133.958 & 0.791 & 2.821$\pm$0.027 & 4.16 & 1.059 & 2.987$\pm$0.029 \\
  3137.933 & 0.903 & 2.564$\pm$0.024 & 3.14 & 1.053 & 2.697$\pm$0.025 \\
  3389.187 & 0.972 & 2.551$\pm$0.037 & 0.99 & 1.049 & 2.676$\pm$0.039 \\
  3405.120 & 0.420 & 3.056$\pm$0.023 & 1.21 & 1.061 & 3.242$\pm$0.024 \\
  3406.117 & 0.448 & 3.037$\pm$0.015 & 0.97 & 1.062 & 3.225$\pm$0.016 \\
  3407.095 & 0.476 & 3.056$\pm$0.025 & 1.05 & 1.062 & 3.245$\pm$0.027 \\
  3408.070 & 0.503 & 3.143$\pm$0.082 & 0.19 & 1.062 & 3.338$\pm$0.087 \\
  3409.144 & 0.533 & 3.071$\pm$0.026 & 0.73 & 1.063 & 3.264$\pm$0.028 \\
  3412.117 & 0.617 & 2.987$\pm$0.030 & 0.44 & 1.063 & 3.175$\pm$0.032 \\
  3413.150 & 0.646 & 2.978$\pm$0.032 & 0.38 & 1.063 & 3.166$\pm$0.034 \\
  3414.140 & 0.674 & 2.859$\pm$0.041 & 0.10 & 1.063 & 3.039$\pm$0.044 \\
  4244.919 & 0.047 & 2.613$\pm$0.020 & 0.51 & 1.049 & 2.741$\pm$0.021 \\
  \hline
  \end{tabular}
\end{table}

\begin{figure}
\begin{center}
  \includegraphics[scale=0.58]{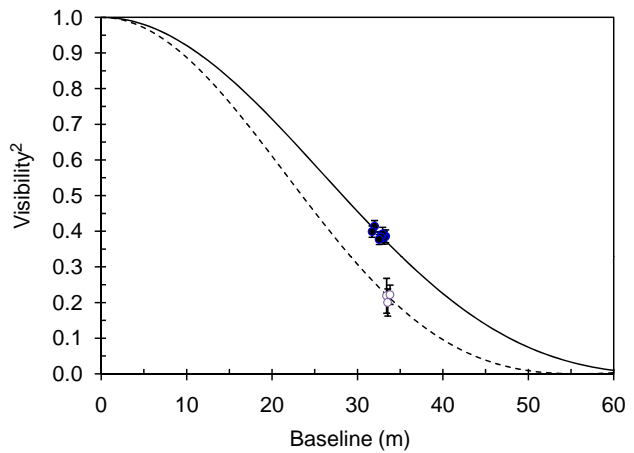}
  \caption{Examples of the data and fitted uniform-disk angular diameter curves for observations
  near minimum and maximum angular diameters.  Key: The filled circles are the data for phase 0.047
  on 24~May~2007 and the solid line is the fit to these data; the open circles are the data for
  phase 0.503 on 6~February~2005 and the dashed line is the fit to these data.  Details of the fitting
  procedure are given in the text.}
  \label{fig:transforms}
\end{center}
\end{figure}

Figure~\ref{fig:transforms} shows the data and fitted uniform-disk angular diameter curves
for near minimum angular diameter (phase = 0.047 on 24 May 2007) and near maximum angular
diameter (phase = 0.503 on 6 February 2005).

\subsection{The Uniform-Disk Angular Diameter of the Calibrator
$\iota$~Car} \label{sec:iotacar}

A check was made on the consistency of the transfer function $T$
determined for the three primary calibrators, namely $\beta$,
$\iota$ and s~Car.  For each pair of consecutive observations of
calibrators the transfer functions were plotted against each
other.  For example, where observations of $\beta$~Car and s~Car
bracketed an observation of $\ell$~Car, the transfer function of
$\beta$~Car was plotted against the transfer function of s~Car.
For all the 40\,m observations this resulted in 67 data points for
s~Car v. $\beta$~Car, 85 points for $\iota$~Car v. $\beta$~Car,
and 51 points for $\iota$~Car v. s~Car.  Linear regression fits of
the form $y = bx$ were made to each of the three plots.

In the case of s~Car v. $\beta$~Car, $b = 1.001\pm0.006$
indicating that these two calibrators are consistent with each
other. However, in the case of $\iota$~Car v. $\beta$~Car, $b =
0.898\pm0.007$, and for $\iota$~Car v. s~Car, $b = 0.897\pm0.009$.
Since both these latter two plots show essentially the same slope,
which differs significantly from the value of unity expected if
the calibrators were consistent, it suggests that the value
adopted for the uniform-disk angular diameter of $\iota$~Car is
too small. One possibility for this inconsistency is that
$\iota$~Car is a binary system but there is no evidence in the
literature to support this.  A careful examination of the SUSI
data for the signature variations expected from a binary system
has proved negative.  It appears that the value adopted for the
uniform-disk angular diameter is too small.  Although it is hard
to justify, arbitrarily increasing the angular diameter of
$\iota$~Car to $1.77\pm0.12$\,mas, $\sim2\sigma$ greater than
predicted, gives $b$ values closest to unity for
$\iota$~Car v. $\beta$~Car and $\iota$~Car v. s~Car
($b = 1.001\pm0.011$ for $\iota$~Car v. s~Car and
$b = 1.003\pm0.009$ for $\iota$~Car v. $\beta$~Car).  The
implications of making this change to the value of the uniform-disk
angular diameter of $\iota$~Car will be discussed in
Section~\ref{sec:discuss} but the analysis will initially be completed
with the data in Table~\ref{tab:ads} which is based on the uniform-disk
angular diameter of $\iota$~Car being equal to $1.55\pm0.12$\,mas.

\subsection{Limb-Darkening Factors} \label{sec:ldfactors}

In order to determine the true, limb-darkened angular diameters,
the uniform-disk angular diameters have to be multiplied by
limb-darkening factors ($\rho_{\lambda}$) that are dependent on
the centre-to-limb (CTL) intensity distributions for the star.  The CTL
intensity distributions are dependent on wavelength, and on the
effective temperature ($T_{\mathrm{eff}}$), surface gravity ($\log{g}$)
and composition ([Fe/H]) of the stellar atmosphere.

The form of the CTL intensity distribution and its phase dependence has been
questioned.  \citet{02marengo, 03marengo} have computed centre-to-limb intensity distributions for
the Cepheid $\zeta$~Gem and shown that their distributions computed for hydrodynamic models in spherical
geometry differ from those for hydrostatic, plane-parallel models, particularly at certain ranges
in phase.  \citet{06narb} have used a hydrodynamic model of $\delta$~Cep to derive intensity
distributions in the continuum and in four spectral lines.  They found that limb-darkening in
the continuum revealed a systematic shift in phase of the derived angular diameter of 0.02.
However, the distance is not affected because it is linked to the amplitude of the angular
diameter curve, which is only slightly changed by the shift effect.  They further claim that
considering the time-dependence of limb-darkening does not seem to be a priority for the IBW
method.  The validity of this claim is considered in Section~\ref{sec:constantLD}
and the use of a phase-dependent limb-darkening factor is justified for the wavelength of the
observations presented here (696\,nm).

In the absence of a CTL intensity distribution computed for a hydrodynamic
model of $\ell$~Car, as done for $\zeta$~Gem, use has been made of the tabulation of
$\rho_{\lambda}$ by \citet{00dtb} computed for the extensive grid of centre-to-limb
intensity variations given by \citet{93ka, 93kb} for his model atmospheres.

\begin{figure}
\begin{center}
  \includegraphics[scale=0.57]{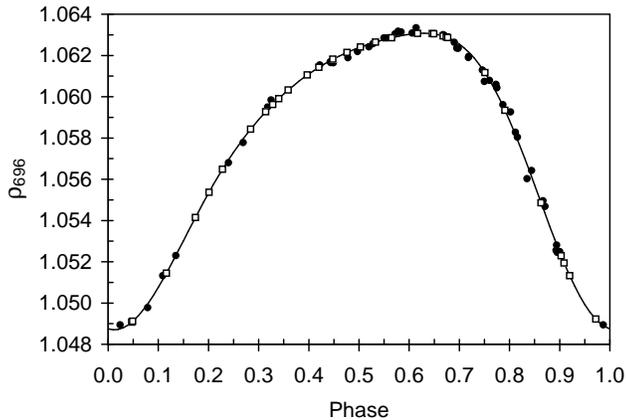}
  \caption{Limb-darkening factor $\rho_{696}$ v. pulsation phase for $\ell$~Car.
  The curve is the sixth order Fourier series fit to the filled circles which are
  the values determined for the phases tabulated by \citet{99mt}.
  The open squares are the values for the epochs of the SUSI determinations
  of the uniform-disk angular diameters, computed from the coefficients of the Fourier fit.
  Further details are given in the text.}
  \label{fig:ld_plot}
\end{center}
\end{figure}

A value of [Fe/H] = 0.3 has been adopted for $\ell$~Car following
\citet{97CdS}.  \citet{99mt} has tabulated 47 values of
$T_{\mathrm{eff}}$ and $\log{g}$ as a function of phase and these
data have been adopted to establish a $\rho_{696}$ versus phase
curve. For each phase tabulated by Taylor a value for $\rho_{696}$
was interpolated from the \citet{00dtb} tabulation. A sixth order
Fourier series was fitted to the resulting values of $\rho_{696}$
versus phase (the lowest order to give a smooth and accurate
representation of the data; the data has a standard deviation
from the fitted curve of 0.00024).  The coefficients of the fit
were then used to compute values for $\rho_{696}$ for the phases
of the SUSI uniform-disk angular diameter determinations. These
values are listed in Table~\ref{tab:ads} and plotted in Figure~\ref{fig:ld_plot}.
Figure~\ref{fig:ld_plot} also includes the values determined for the Taylor phases and the curve fitted to them.
Taylor included estimates of the uncertainties in the values of $T_{\mathrm{eff}}$
and $\log{g}$ in her tabulation and from these uncertainties it is
estimated that the values of $\rho_{696}$ in Table~\ref{tab:ads}
are accurate to $\pm0.001$, with the caveat that we are assuming that
the Kurucz models upon which \citet{00dtb} based their tabulation of limb-darkening
factors are appropriate for $\ell$~Car.  The mean value of $\rho_{696}$
for the fitted curve is 1.057.

\subsection{The Limb-Darkened Angular Diameters} \label{sec:ldads}

The uniform-disk angular diameters have been converted to
limb-darkened angular diameters ($\theta_{\mathrm{LD}}$) in
Table~\ref{tab:ads} using the listed limb-darkening correction
factors.  The uncertainties in the uniform-disk and limb-darkened
angular diameters given in Table~\ref{tab:ads} do not include the
uncertainty in the effective wavelength, which is a systematic
uncertainty of $\pm$0.3\%.  Its effect on the mean angular diameter
and on the distance to $\ell$~Car will be discussed in
Section~\ref{sec:discuss}.  The uncertainty in the limb-darkened angular
diameter in Table~\ref{tab:ads} does however include the uncertainty in
the limb-darkening correction factor.

\section{The Spectroscopic Data} \label{sec:spec}

In order to determine the distance and mean diameter of $\ell$~Car
the radial displacements of the stellar surface as a function of
phase are required for combination with the limb-darkened angular
diameters. The starting point for establishing the radial
displacements is the observed radial velocity curve as a function
of phase.

\subsection{The Radial Velocity Data}

The most extensive radial velocity data set is that by \citet{97tetal}
based on 67 spectra from Mount Stromlo Observatory (MSO) in Australia and
70 spectra from Mount John University Observatory (MJUO) in New Zealand.
Early measurements of radial velocity induced spectral line displacements were
generally made by line centroid or bisection estimates made by eye
(e.g. \citealt{69dawe}) which was the last published velocity curve for
$\ell$~Car prior to \citet{97tetal}).

\citet{97tetal} determined radial velocities by averaging the radial
velocities for 19 metallic lines measured by the line-bisector method
\citep{92waller} from an average of depths 0.7, 0.8 and 0.9 (continuum at
0.0 and core at 1.0).  The MSO and MJUO sets of data were initially phased using
the ephemeris of \citet{92shob} but \citet{97tetal} found a small phase shift
between the earliest MJUO observations and later observations and derived a
small phase correction to bring all the MSO and MJUO observations into phase
alignment.  We have adopted their tabulated data that include the phase adjustments
with three exceptions.  The Julian Dates 2449408.65 and 2450380.57 for MSO data
and 2449408.66 for MJUO data correspond to times during daylight hours so data
for these dates have been omitted leaving a total of 134 data points.

\citet{02ber} lists HJDs for 19 radial velocity measurements for $\ell$~Car but
does not give phases.  For $\ell$~Car the maximum difference between HJD and JD is
negligible at less than $3\times10^{-5}$ of the pulsation period and has been ignored.
Phases have been computed with Shobbrook's ephemeris and the
resulting data are in good phase agreement with the data of \citet{97tetal},
discussed above, but there is an offset in radial velocity.  \citet{04K2}
first noted this and chose `to shift the \citet{97tetal} data set by $-1.5$\,km s$^{-1}$
to bring all the data on the well established CORAVEL system of \citet{02ber}'.  Since
the data set of \citet{97tetal} is significantly more numerous than that of \citet{02ber}
(134 data points versus 19), and the decision has been made to adopt the \citet{97tetal}
as the basic data set, we have adjusted the Bersier radial velocities.  This decision is
supported by the work of \citet{98kiss} who found that, while `CORAVEL measurements have
excellent internal accuracy, their absolute values are very uncertain'.  The Bersier data,
determined by the cross-correlation technique rather than the line-bisector method employed by
\citet{97tetal}, were combined with the Taylor et al. data with a range of offsets for the Bersier radial velocities.
For each offset a sixth order Fourier series was fitted over the range in phase from -0.037 to 0.75
(the reasons for the choice of the order of the fit and of the range in phase will be discussed in
Section~\ref{sec:rvcurve}) and the minimum value of $\chi^{2}$ corresponds to the addition
of 2.0\,km s$^{-1}$ to the Bersier data.  In view of the resulting good agreement between
the Bersier and Taylor et al. data over the whole pulsation cycle including the amplitude
of the radial velocity variation, the Bersier data have been accepted in spite of the
different technique used for measuring the line shifts.

\citet{05pet} have measured 34 radial velocities at the MJUO and while these are more
recent than those of \citet{97tetal}, the technique for measuring the radial velocities
by the line-bisector method is identical.  As well as giving the JDs of the observations the
authors have listed phases based on an ephemeris by \citet{76pel} which are significantly offset
from the data of Bersier and Taylor et al.  We have computed the phases from the given JDs,
using the ephemeris of \citet{92shob}, and find a small phase offset from the Taylor et al.
data suggesting a glitch or period change between the Taylor et al. and the later
Petterson et al. observations.  In order to establish the optimum phase adjustment the
Petterson et al. data were combined with the Taylor et al. data with a range of phase offsets.
For each offset a sixth order Fourier series was fitted over the range in phase from -0.037 to 0.75,
as in the case of the Bersier radial velocities, and the minimum value of $\chi^{2}$ corresponds
to the subtraction of 0.02 in phase from the \citet{05pet} data.  The adjusted Petterson et al. data
are in excellent agreement with the combined Taylor et al. and Bersier radial velocity data with
the exception of the five data points for JD 2451163.0392, 2450683.1543, 2450683.8039,
2450684.1304 and 2450684.8105 which all lie several times the quoted uncertainties from the combined
Taylor et al., Bersier, and Petterson et al. plot of radial velocity against phase.  The Petterson et al.
data have been accepted with the exception of the five discrepant points.

\citet{06nara} have published radial velocities for $\ell$~Car determined by three different methods
of measuring line displacements.  These methods differ from those used for the data considered so far and result
in a range of amplitudes of the radial velocity curve bracketing the amplitude of the curve determined
from the Taylor et al., Bersier and Petterson et al. curve with none of the three agreeing with it.  In view
of the disagreement with the other three sources the \citet{06nara} data have not been included.

\subsection{The Radial Velocity Curve} \label{sec:rvcurve}

\begin{figure}
\begin{center}
  \includegraphics[scale=0.50]{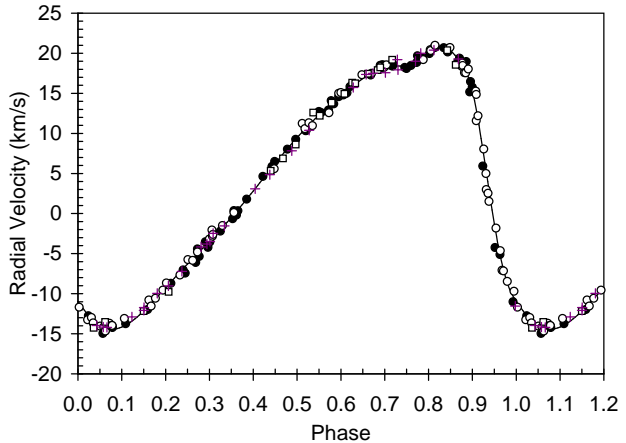}
  \caption{The assembled radial velocity curve v. phase for
  $\ell$~Car.  The line is the fit to the data; filled circles are
  MSO data from \citet{97tetal}; open circles are MJUO data from
  \citet{97tetal}; open squares are data by \citet{02ber}; and
  crosses are data by \citet{05pet}.  Details are given in the
  text.}
  \label{fig:radvels}
\end{center}
\end{figure}

Figure~\ref{fig:radvels} shows the assembled radial velocity data versus phase together with the fitted curve.
Attempts were made to fit a Fourier series to the assembled data but even with a sixteenth order series the
fit was poor in parts.  \citet{99mt} overcame this difficulty by fitting a sixth order Fourier fit to the ascending
branch of the curve and a ninth order fit to the descending branch.  After some experimentation it was found that
the good fit shown in Figure~\ref{fig:radvels} could be obtained by dividing the curve into three ranges in phase
and making a separate sixth order Fourier series fit to each.  The ranges of the fits, which overlap, are
listed in Table~\ref{tab:rvfits}.  In each case the reduced $\chi^{2}$ values indicated that the published uncertainties
in the radial velocities are optimistic.  Adopting uncertainties of $\pm$0.45\,kms$^{-1}$, $\pm$0.50\,kms$^{-1}$, and
$\pm$1.0\,kms$^{-1}$ for the rising, middle, and falling sections respectively gave reduced $\chi^{2}$ values close to
unity.  The coefficients of the fits were employed to compute radial velocities at
intervals of 0.005 in phase over the ranges of the fits and plots were made of the fits in the overlapping regions.
The phases for the transitions from one fit to the next were chosen by inspection of these plots and were the phases
at which the overlapping fits were closest to each other.  For the transition between the rising and middle sections
the radial velocities at phase 0.710 were adjusted to an intermediate value to give a smooth transition between the
fits on either side.  Similarly, for the transition between the middle and falling sections, the radial velocity at
phase 0.890 was adjusted and, for the transition between the falling and rising sections, the radial velocity at
phase 0.015 was adjusted.  The adjustments to the radial velocities were small, averaging $\sim$0.3\% of the mean
radial velocities at the transitions.  The fits are summarised and the phases at which they
were matched are listed in Table~\ref{tab:rvfits}.

\begin{table}
  \caption{Details of the Fourier series fits to the selected radial velocity data in
  three ranges in phase.  N is the number of data points in each range in phase, N$_{\mathrm{F}}$
  is the order of the Fourier series fit, and Phase(M) is the phase at which the separate fits were
  matched.  Further details are given in the text.}
  \label{tab:rvfits}
  \begin{tabular}{ccccc}
  \hline
   Section & Phase & N &  N$_{\mathrm{F}}$ & Phase(M) \\
    & Range &  &  &  \\
  \hline
  Rising &-0.037 to 0.75 & 137 & 6 & \\
  Rising to Middle & & & & 0.710 \\
  Middle & 0.701 to 0.896 & 42 & 6 & \\
  Middle to Falling & & & & 0.890 \\
  Falling & 0.862 to 1.031 & 38 & 6 & \\
  Falling to Rising & & & & 0.015 \\
  \hline
  \end{tabular}
\end{table}

The radial velocity curve shown in Figure~\ref{fig:radvels}, assembled from the Fourier series
fits listed in Table~\ref{tab:rvfits}, has been adopted for the subsequent analysis

\subsection{The $p$-Factor} \label{sec:p-factor}

The radial velocity curve in Figure~\ref{fig:radvels} is for the measured radial velocities but
does not represent the true radial velocity of the stellar surface since it is a value integrated
over the stellar surface.  It includes projection and limb-darkening effects that vary from the
centre to the limb of the star.  These effects depend on spectral line shape, which is not only a
function of phase, but is also affected by the velocity structure within the line-forming region and
the contributions from the different layers of the atmosphere.  Conventionally, in the absence of
sufficiently detailed modelling of the stellar atmosphere, the measured radial velocities are
multiplied by a constant, known as the $p$-factor, to correct them to the radial velocity of the
stellar surface.  There are a number of issues to be considered in the choice of $p$ since any
error or uncertainty in the value adopted translates directly to the distance determined for the
Cepheid when the spectroscopic and interferometric data are combined.

The evaluation of $p$-factors is generally derived from model stellar atmosphere models but
\citet{05merand} have presented a measured value of 1.27$\pm$0.06 for the $p$-factor for $\delta$~Cep
based on interferometric measurements with the CHARA Array and on the Hubble Space Telescope parallax \citep{02ben}.
The accuracy is limited by the parallax and the authors conclude that theoretical studies using
realistic hydrodynamical codes are needed.

The literature dealing with the theoretical evaluation of the $p$-factor from model atmosphere studies
is extensive but with no clear results applicable to $\ell$~Car.  While \citet{95sabbey} claim that the phase
dependence of $p$ increases the Baade-Wesselink (BW) radius by $\sim4-6$\%, depending on the
constant value of  $p$ used for comparison, \citet{04nar} claim that their choice of a constant
$p$-factor for the Interferometric Baade-Wesselink method (IBW), compared to a time-dependent one,
leads to a systematic error of the order of only 0.2\% in the final distance determination for $\delta$~Cep.

The projection factor is also sensitive to the centre-to-limb intensity distribution or
limb darkening, but it is not clear whether a mean value for the limb-darkening is adequate or whether
a phase dependence is significant for the p-factor.  However, it is clear that the phase dependence of
limb darkening is significant for the conversion of uniform-disk angular diameters to limb-darkened
angular diameters at 696\,nm, as mentioned in Section~\ref{sec:ldfactors} and justified in
Section~\ref{sec:constantLD}.

In most cases more detailed modelling and, in particular, the hydrodynamic modelling in
spherical geometry that predicts a phase dependence of $p$ has been done for a particular Cepheid, most
commonly $\delta$~Cep or $\zeta$~Gem, and the results are not in a form that can be scaled to other
Cepheids.  The concluding recommendation of these studies is generally that each Cepheid should be
individually modelled.  This is not within the scope of our programme and we have therefore decided
to use a fixed value but to make available all the relevant data to enable the results to be updated
by others when improved values for $p$, or an appropriate model for $\ell$~Car have been developed.

A brief summary and discussion of the fixed values of $p$ adopted for $\ell$~Car in the literature
is appropriate at this point.  As noted by \citet{97tetal} the value of $p$ depends not only on the
technique used to measure the radial velocities, but also upon the strength of the lines chosen and
their wavelengths.  \citet{97tetal} adopted a value of 1.38$\pm$0.03 \citep{94albrow} but revised
their results \citep{98tandb} using $p = 1.39-0.03\log{P} = 1.34$ (\citealt{86handb}; \citealt*{93gie}).
\citet{04K2}, in a comparison of the interferometric and surface brightness techniques, used the same
formula with $p = 1.343$ for $\ell$~Car.  In this comparison they only considered radial velocity
points in the phase interval 0.0 to 0.8 following \citet{04storm}.  However, \citet{04K1} adopted a
value of 1.36 for $p$ for all the 7 Cepheids in their programme which included $\ell$~Car and
justified it on the grounds that \citet*{82bur} had shown that this value was appropriate for the
radial velocity measurements by \cite{02ber} that they had used.  \citet{07nar} have
used a hydrodynamic model of $\ell$~Car to validate a spectroscopic method of determining the
$p$-factor in which it was divided into three sub-concepts.  While their work is not directly
applicable to our data, because it was restricted to a single specific spectral line and the
method employed for measuring the line differed from that for the data we are using, they derived a
value of 1.27$\pm$0.02 for $p$.  \citet{07groen} has evaluated a relationship between the $p$-factor
and pulsation period based on five Cepheids with interferometrically measured angular diameter variations and
known distances taken from the literature.  Based on a total of seven stars with periods in the range
5-35 days it is claimed that there is no evidence for a period dependence of the $p$-factor although values found
range from 1.193 to 1.706, albeit with large uncertainties in the majority of cases.  For $\ell$~Car the
$p$-factor is found to be 1.193$\pm$0.058$\pm$0.120 based on the distance of 498$\pm$50\,pc determined by
\citet{07ben} with the Hubble Space Telescope Fine Guidance Sensors.  The first uncertainty is from the
fitting process and the second is due to the uncertainty in the distance.  The majority of distance determinations
to $\ell$~Car to date, which are listed in Table~\ref{tab:distance} and include the distance determined here, suggest a
larger value for the distance to $\ell$~Car, implying a larger value for the $p$-factor of the order of 1.29
if determined by the approach employed by Groenewegen.  Based on the five Cepheids analysed, which had
interferometrically determined angular diameters and distances, \citet{07groen} concluded that a constant value of
$p = 1.25\pm0.05$ was appropriate.  But, as has been discussed, at least in the case of $\ell$~Car a larger value
is indicated.

To summarise, there is no agreement on the optimum fixed value of the $p$-factor to use although it
is clear that it depends on the individual Cepheid and on the details of the observing and analysis
techniques used for both spectroscopy and interferometry.  Faced with these difficulties we have
adopted a constant value of 1.30$\pm$0.05.  The influence of this decision on the
mean diameter and distance of $\ell$~Car will be discussed in Section~\ref{sec:discuss}.

\subsection{The Radial Displacement Curve} \label{sec:rdcurve}

Integration of the radial velocity curve shown in Figure~\ref{fig:radvels} gave the radial velocity
of the centre of mass $V_{\gamma}$ equal to $+4.13\pm$0.01\,kms$^{-1}$ where the uncertainty has been
estimated using the bootstrapping method.  This differs from the value of $+4.21\pm$0.01\,kms$^{-1}$
given by \citet{97tetal} but is based on the assembly of a larger body of radial velocity data.

The radial displacement of the stellar surface in solar radii $\Delta R$($\phi$), as a function of
phase $\phi$, was found by integrating the radial velocity $V_{\mathrm{RV}}(\phi)$,
after correction for $V_{\gamma}$, using

\begin{equation}
\Delta R(\phi) = -\frac{pP}{R_{\sun}}\int(V_{\mathrm{RV}}(\phi) - V_{\gamma})\mathrm{d}\phi
\end{equation}
where $P$ is the pulsation period in seconds and $R_{\sun}$ is the solar
radius in km.

The integrations were made at intervals in phase of 0.005 and the resulting values of radial
displacement, after the determination and subtraction of the mean value, were fitted with a
Fourier series.  Series of increasing order were fitted and it was found that the reduced $\chi^{2}$ value
for the fits had a minimum value for a sixteenth order fit.  This has been adopted for the
subsequent analysis.  The uncertainty in the radial displacements has been evaluated following
\citet{97tetal} using the expression by \citet{77bal}:

\begin{equation}
\sigma(\mathrm{RD}) = \frac{pP\sigma(\mathrm{RV})}{2R_{\sun}\surd\overline{N}} \label{eqn:sigmard}
\end{equation}
where $\sigma$(RD) is the uncertainty in radial displacement in solar radii, $P$ is
the pulsation period in seconds, $\sigma$(RV) is the standard deviation of the radial
velocities about the fitted curve ($\sigma$(RV) = 0.6\,km s$^{-1}$), $R_{\sun}$ is the
solar radius in km, and N is the number of observations.  Substitution in
equation~(\ref{eqn:sigmard}) gives $\sigma$(RD) = 0.13\,$R_{\sun}$ (less than 0.4\% of the total radial
displacement due to the Cepheid pulsation).

\section{The Combination of Interferometric and Spectroscopic
Data} \label{sec:ldadvrd}

The relationship between the limb-darkened angular diameter
$\theta_{\mathrm{LDobs}}$ and the radial displacement of the
Cepheid surface is given by

\begin{equation}
\theta_{\mathrm{LDobs}}(\phi_{i}) =
\overline{\theta_{\mathrm{LD}}} + 9.298\left(\frac{\Delta
R(\phi_{i})}{d}\right) \hspace{5mm} \mbox{mas} \label{eq:fit}
\end{equation}
where $\overline{\theta_{\mathrm{LD}}}$ is the mean limb-darkened angular diameter (the
limb-darkened angular diameter at zero displacement),
$\Delta R(\phi_{i})$ is the radial displacement in solar radii for the
$i$th observation at phase $\phi_{i}$, $d$ is the distance in pc, and the constant
converts the term in parentheses into mas using values for the solar radius and parsec
in metres from \citet{cox}.

\begin{figure}
\begin{center}
  \includegraphics[scale=0.60]{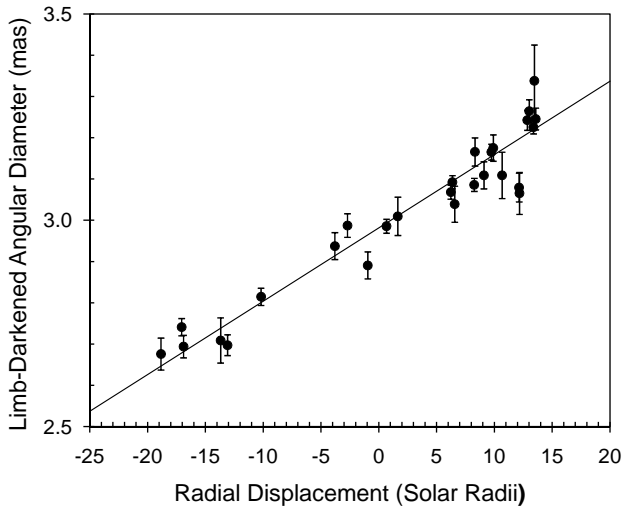}
  \caption{The limb-darkened angular diameter versus radial displacement for
  $\ell$~Car.  The filled circles with error bars are the observed
  values of limb-darkened angular diameter from Table~\ref{tab:ads}
  and the line is the fit to angular diameter versus radial displacement
  using equation~(\ref{eq:fit}).  Further details are given in the text.}
  \label{fig:ldadvrd}
\end{center}
\end{figure}

A weighted linear least squares fit has been made to $\theta_{\mathrm{LD}}(\phi_{i})$
versus $\Delta R(\phi_{i})$ to determine $\overline{\theta_{\mathrm{LD}}}$ and
$d$ for $\ell$\,Car.   A small phase shift was found between the angular diameter
and radial velocity data, which is not surprising since the epoch of the first SUSI
observation was more than 40 pulsation periods after the last radial velocity
observation.  The phase offset was established by repeating the fit with a range
of phase offsets to find the minimum value of reduced $\chi^{2}$ for the fit.
The phase offset was found to be a correction of -0.0635$\pm$0.015 to the phases
of the radial displacements relative to the phases of the angular diameters.
The fit using this offset gives the mean limb-darkened angular diameter of $\ell$~Car equal
to 2.981$\pm$0.005\,mas and the distance to $\ell$~Car equal to 523$\pm$15\,pc.
The observational data are shown in Figure~\ref{fig:ldadvrd} with the fitted line.

In Section~\ref{sec:iotacar} the apparent inconsistency of the uniform-disk
angular diameter of $\iota$~Car was discussed and the possibility of increasing
it to 1.77$\pm$0.12\,mas was proposed although all the analysis to this point
has been carried out with our estimate for its angular diameter of 1.55$\pm$0.12\,mas.
In order to investigate the effect of increasing the angular diameter the analysis
to this point has been repeated with the angular diameter of $\iota$~Car equal to
1.77$\pm$0.12\,mas.  The results are: phase offset -0.064$\pm$0.001, mean limb-darkened
angular diameter of $\ell$~Car equal to 2.999$\pm$0.005\,mas and the distance to
$\ell$~Car equal to 526$\pm$15\,pc.

At this point it is stressed that the uncertainties are those given by the
fits.  However, it can be seen that the change in the angular diameter of $\iota$~Car
has essentially negligible effect on the distance determination compared with the uncertainty
in the distance, but does increase the mean angular diameter by more than three times the formal
uncertainty from the fits.  The difference in the mean angular diameter and systematic
uncertainties that have not been considered so far are discussed in Section~\ref{sec:discuss}.

\begin{figure}
\begin{center}
  \includegraphics[scale=0.50]{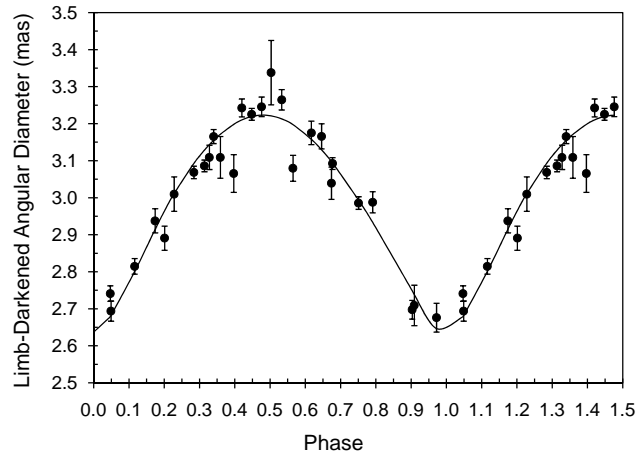}
  \caption{The limb-darkened angular diameter versus phase for
  $\ell$~Car.  The filled circles with error bars are the observed
  values of limb-darkened angular diameter from Table~\ref{tab:ads}
  and the line is the computed limb-darkened angular diameter versus phase
  from the fit to angular diameter versus radial displacement
  using equation~(\ref{eq:fit}).  The data points and curve in the
  phase interval 0--0.5 have been repeated for the phase interval 1.0--1.5.
  Further details are given in the text.}
  \label{fig:advphase}
\end{center}
\end{figure}

The limb-darkened angular diameters versus phase for $\ell$~Car, assuming the angular
diameter of $\iota$~Car equal to 1.55$\pm$0.12\,mas, are shown in Figure~\ref{fig:advphase}
together with the limb-darkened angular diameter versus phase curve computed for the fit of
limb-darkened angular diameter versus radial displacement.

\section{Discussion}\label{sec:discuss}

In Section~\ref{sec:ldadvrd} two values were presented for the mean limb-darkened
angular diameter of $\ell$~Car: 2.981$\pm$0.005\,mas for the uniform-disk angular
diameter of the calibrator $\iota$~Car equal to 1.55\,mas and 2.999$\pm$0.005\,mas
for $\iota$~Car equal to 1.77\,mas.  Increasing the uniform-disk angular diameter
of $\iota$~Car produced greater consistency between the three calibrators, as discussed
in Section~\ref{sec:iotacar}, but the change was arbitrary and we cannot be certain that
the larger value is correct.  We therefore adopt the mean of the two results and assign
an uncertainty to cover the uncertainties of the two values.  In addition there is a
systematic uncertainty due to the uncertainty in the effective wavelength of $\pm$0.3\,per cent.
Thus the final value for the mean limb-darkened angular diameter of $\ell$~Car, with the
systematic uncertainty in parentheses, is
$\overline{\theta_{\mathrm{LD}}}$ = 2.990$\pm$0.014 ($\pm$0.009)\,mas.  This value is in
excellent agreement with the only other direct interferometric determination by \cite{04K1}
of 2.988$\pm$0.012\,mas.

The corresponding two values for the distance presented in Section~\ref{sec:ldadvrd}, namely
523$\pm$15\,pc and 526$\pm$15\,pc, are in good agreement and we adopt 525$\pm$16\,pc for the
distance.  There are two systematic uncertainties in the distance due to the effective wavelength ($\pm$0.3\,per cent)
and the $p$-factor ($\pm$3.8\,per cent).  The value for the distance to $\ell$~Car, with the
systematic uncertainty in parentheses, is 525$\pm$16 ($\pm$20)\,pc.  Combining the uncertainties quadratically
gives the distance as 525$\pm$26\,pc.

\subsection{A Constant versus a Phase-Dependent Limb-Darkening Factor} \label{sec:constantLD}

The limb-darkening factor for converting uniform-disk angular diameters to
limb-darkened angular diameters is expected to vary with pulsation phase and we
have taken this into account as discussed in Section~\ref{sec:ldfactors}.  However,
the question has been asked as to whether this was necessary and would a constant
value, equal to the mean value taken from the curve in Figure~\ref{fig:ld_plot}, have
given different results.  To examine this, the entire analysis has been repeated using
the mean value of the limb-darkening factor of 1.057, given in Section~\ref{sec:ldfactors},
in place of the phase-dependent values given in Table~\ref{tab:ads}.

The mean limb-darkened angular diameter was found to be unchanged as expected. The mean
limb-darkened angular diameter of $\ell$~Car was 2.979$\pm$0.017\,mas for the uniform-disk angular
diameter of the calibrator $\iota$~Car equal to 1.55\,mas and 2.997$\pm$0.017\,mas
for $\iota$~Car equal to 1.77\,mas, where the statistical and systematic uncertainties have
been combined quadratically.  In each case the values are slightly smaller than the values
for the phase-dependent limb-darkening factor (0.002\,mas) but this is a result of the
rounding of the mean value of the limb-darkening factor.

The distance to $\ell$~Car is changed because the mean limb-darkening factor results in
a smaller angular pulsation amplitude (0.525$\pm$0.018\,mas compared with 0.560$\pm$0.018\,mas).
The corresponding values for the distance are 559$\pm$17\,pc and
561$\pm$17\,pc compared with 523$\pm$15\,pc and 526$\pm$15\,pc for the phase-dependent limb-darkening factor.
The mean of the two values in each case is 560$\pm$18\,pc and 525$\pm$16\,pc.  The difference is significant
and justifies the use of the phase-dependent limb-darkening factor.   \citet{04K2} neglected
the phase dependence of the limb-darkening factor in their study of $\ell$~Car.  Their decision was based
on the estimate by \citet{03marengo} that the variation would be less than 0.3\,per cent peak to peak in
the $H$ band for $\zeta$~Gem and the fact that it would be even less in the $K$ band.  At 696\,nm the limb-darkening
factor varies by more than 1.3\,per cent, as shown in Figure~\ref{fig:ld_plot}, and cannot be ignored.

\begin{table*}
  \caption{The distance to $\ell$~Car.  $p$ is the $p$-factor, the column headed ID contains
  acronyms for the methods employed with the key at the foot of the table, and the key to the
  numbered references is also at the foot of the table.  Further details are given in the text.}
  \label{tab:distance}
  \begin{tabular}{llllllc}
  \hline
   \multicolumn{1}{c}{Distance} & \multicolumn{1}{c}{Radius} & \multicolumn{1}{c}{$p$} & \multicolumn{1}{c}{ID} &  \multicolumn{1}{c}{Distance} & \multicolumn{1}{c}{Radius} & Reference \\
    & & &  & \multicolumn{1}{c}{for $p$ = 1.30} & \multicolumn{1}{c}{for $p$ = 1.30} & \\
   \multicolumn{1}{c}{(pc)} & \multicolumn{1}{c}{($R_{\sun}$)} & & & \multicolumn{1}{c}{(pc)} & \multicolumn{1}{c}{($R_{\sun}$)} & \\
  \hline
  550$\pm$17 & 173$\pm$5 & 1.34 & BE & 534$\pm$22  & 168$\pm$7 & 1 \\
  $566^{+24}_{-19}$ & $182^{+8}_{-7}$ & 1.343 & IBW & $548^{+30}_{-27}$ & $176^{+10}_{-9}$ & 2 \\
  560$\pm$23 & 179$\pm$7 & 1.343 & IRSB & 542$\pm$30  & 173$\pm$10 & 2 \\
  559$\pm$19 & 179.9$\pm$6.4 & 1.343 & IRSB-Bay & 541$\pm$27  & 174$\pm$9 & 3 \\
  485$\pm$64 & \multicolumn{1}{c}{---} & \multicolumn{1}{c}{---} & RH  &  \multicolumn{1}{c}{---} & \multicolumn{1}{c}{---} & 4 \\
  498$\pm$50 & \multicolumn{1}{c}{---} & \multicolumn{1}{c}{---} & HST &  \multicolumn{1}{c}{---} & \multicolumn{1}{c}{---} & 5 \\
  525$\pm$26 & 168.8$\pm$8.2 & 1.30 & IBW & 525$\pm$26 & 169$\pm$8 & This work \\
  \hline
  \end{tabular}
\newline
Key to ID acronyms: BE - Barnes-Evans; IBW - Interferometric Baade-Wesselink;
IRSB - Infrared Surface Brightness; IRSB-Bay - Infrared Surface Brightness (Bayesian);
RH - Revised Hipparcos Parallax; HST - Hubble Space Telescope Fine Guidance Sensors.
\newline
Key to References: 1 - \citet{98tandb}; 2 - \citet{04K2}; 3 - \citet{05betal};
4 - \citet{07vanL}; 5 - \citet{07ben}.
\end{table*}

\subsection{A Comparison of Distance and Radius Determinations} \label{sec:comparison}

Table~\ref{tab:distance} lists recent values for the distance to $\ell$~Car from the literature
together with the value determined in this work for a phase-dependent limb-darkening factor.
For the latter the statistical and systematic
uncertainties have been combined quadratically.  The value by \citet{98tandb} using the BE method,
which succeeds the value by \citet{97tetal}, has an unrealistic published uncertainty of $\pm$4\,pc because
systematic effects have not been taken into account.  \citet*{97gie} have shown that a systematic uncertainty
of the order of $\pm$3\,per cent should be taken into account for the surface brightness method and the uncertainty
for the \citet{97tetal} distance in Table~\ref{tab:distance} has been increased to include this additional
uncertainty.   It is noted that the values by \citet{04K2} using the IBW and IRSB methods, and by \citet{05betal}
by the IRSB method, do not include an allowance for the systematic uncertainty in the $p$-factor adopted.

The distance values in Table~\ref{tab:distance} that involve the $p$-factor have used different values to that
adopted here and, while it is recognised that different values \textit{may} be appropriate for different methods, a
column has been added to the table with the distances scaled to a common value of $p$ = 1.30$\pm$0.05.  The uncertainties
associated with this column have been increased to combine quadratically the uncertainty in the $p$-factor with the
published uncertainty where appropriate.  The uncertainty in the value by \citet{97tetal} has been increased further
as \citet{97gie} included a smaller uncertainty in the $p$-factor ($\pm$2.5\%) in their evaluation of the systematic
uncertainty.  Examination of the distances for the common $p$-factor show excellent agreement for all values.

Columns for the mean radius of $\ell$~Car are included in Table~\ref{tab:distance}, one for the published values and
one for a common $p$-factor, following the same approach as for the distance.  The uncertainty in the limb-darkened
angular diameter for the IBW determinations has been taken into account although it is negligible compared with
the uncertainty in the distance.  Examination of the mean radii for the common $p$-factor also show excellent
agreement for all values.

The mean total range in limb-darkened angular diameter has been taken from the curves fitted to limb-darkened
angular diameter versus radial displacement for the two values of uniform-disk angular diameter for the
calibrator $\iota$~Car.  It is 0.560$\pm$0.018\,mas, which corresponds to 18.7$\pm0.6$\,per cent of the mean stellar diameter.

\section{Conclusion}\label{sec:summary}

The measurement of the angular pulsations of Cepheid variables and the combination with spectroscopically
determined radial displacements of the stellar atmospheres was one of the key programmes for which SUSI
was developed. Only recently has adequate sensitivity been achieved for this programme to be undertaken and
the first results are those for $\ell$~Car presented here.

The angular pulsation of $\ell$~Car has been measured with good cover in pulsation phase and careful attention
has been paid to uncertainties in the measurements, in the adopted calibrator angular diameters, in the
projected value of $V^{2}$ at zero baseline, and to systematic effects.  A phase-dependent limb-darkening
factor, to convert uniform-disk angular diameters to limb-darkened angular diameters, was established based
on Kurucz model atmospheres and its use, rather than a fixed mean value for the limb-darkening factor, was justified for 696\,nm.
The resulting mean limb-darkened angular diameter is 2.990$\pm$0.017\,mas (i.e. $\pm$0.6\,per cent) with a
maximum-to-minimum amplitude of 0.560$\pm$0.018\,mas corresponding to 18.7$\pm$0.6\,per cent in the mean stellar
diameter.  The projected value of $V^{2}$ at zero baseline shows no evidence at 696\,nm of the circumstellar
envelope observed in the $N$ and $K$ bands by \citet{06ketal}.

A radial displacement curve has been computed from carefully selected radial velocity measurements from the
literature and this has been combined with the interferometric data to determine the mean radius and distance
to $\ell$~Car.  The value for the distance is 525$\pm$26\,pc and the mean radius is 169$\pm$8\,$R_{\sun}$.  These
values have been compared with values in the literature and excellent agreement is found, particularly when
all values are reduced to a common $p$-factor.

The uncertainty in the $p$-factor is a major contributor to the uncertainties in the distance and mean radius.
The SUSI data may be useful in the future if $\ell$~Car specific hydrodynamic spherical models are generated that
consider the possibility of a phase-dependence of limb-darkening and the $p$-factor as discussed, for example, by
\citet{94sass} and \citet{02marengo}.  To facilitate such possible applications the calibrated $V^{2}$ values
may be obtained from the lead author (JD).

\section{Acknowledgements}

The SUSI programme is funded jointly by the Australian Research
Council and the University of Sydney.  MJI acknowledges the support of an Australian
Postgraduate Award, APJ and JRN the support of University of Sydney
Postgraduate Awards and APJ the support of a Denison Postgraduate Award during
the course of this work.  We are grateful to the referee who made a number of suggestions that have
helped clarify and justify the approach we have adopted in our analysis.  This research has made
use of the SIMBAD data base, operated at CDS, Strasbourg, France.

\bsp

\end{document}